\documentclass[twocolumn]{aa}
\usepackage{graphicx}
\usepackage{txfonts}
\usepackage{psfig}

\newcommand{\be}{\begin{equation}}
\newcommand{\ee}{\end{equation}}
\def\ltsima{$\; \buildrel < \over \sim \;$}
\def\lsim{\lower.5ex\hbox{\ltsima}}
\def\gtsima{$\; \buildrel > \over \sim \;$}\def\gsim{\lower.5ex\hbox{\gtsima}}

\begin{document}

\title{A refined position catalogue of the {\it Swift} XRT afterglows}
\author{A. Moretti\inst{1}, M. Perri\inst{2}, M. Capalbi\inst{2}, L. Angelini \inst{3,4}, 
J.E. Hill\inst{4,5}, S. Campana\inst{1}, D.N. Burrows\inst{6}, J.P. Osborne\inst{7}, 
G. Tagliaferri\inst{1}, G. Cusumano\inst{8}, P. Giommi\inst{2},
P. Romano\inst{1}, T. Mineo\inst{8}, J. Kennea\inst{6}, D. Morris\inst{6}, J. Nousek\inst{6}, 
C. Pagani\inst{1,6}, J. Racusin\inst{6}, A.F. Abbey\inst{7}, A.P. Beardmore\inst{7}, O. Godet\inst{7}, 
M. R. Goad\inst{7}, K.L. Page\inst{7}, A.A. Wells\inst{7}, G. Chincarini\inst{1,9}}
\offprints{moretti@merate.mi.astro.it}
\institute{
INAF, Osservatorio Astronomico di Brera, Via E. Bianchi 46, I-23807, Merate (LC), Italy
\and
ASI Science Data Center, via G.\ Galilei, I-00044 Frascati, Italy
\and 
NASA/Goddard Space Flight Center, Greenbelt Road, Greenbelt, MD20771, USA
\and       
Department of Physics and Astronomy, Johns Hopkins University, 3400 Charles Street, Baltimore, MD 21218, USA
\and 
Universities Space Research Association, 10211 Wincopin Circle, Suite 500, Columbia, MD, 21044-3432, USA
\and
Department of Astronomy \& Astrophysics, Pennsylvania State University, 525 Davey Lab, University Park, PA 16802, USA
\and
X-Ray Observational Astronomy Group, Department of Physics \& Astronomy, University of Leicester, LE1 7RH, UK
\and
INAF, Istituto di Astrofisica Spaziale e Fisica Cosmica Sezione di Palermo, Via U.\ La Malfa 153, I-90146 Palermo, Italy
\and
Universit\`a degli Studi di Milano-Bicocca, Dipartimento di Fisica, Piazza delle Scienze 3, I-20126 Milano, Italy 
}
\date{Received ; accepted }
\titlerunning{Refined X--ray positions}
\authorrunning{Moretti et al.\ }
\abstract{
We present a catalogue of refined positions of 68 gamma ray burst (GRB) afterglows observed by the 
{\it Swift} X--ray Telescope (XRT) from the launch up to 2005 Oct 16. 
This is a result of the refinement of the  XRT boresight calibration.
We tested this correction by means of a systematic study of a large sample of X--ray sources observed by XRT
with well established optical counterparts.
We found that we can reduce the systematic error radius of the measurements by 
a factor of two, from 6.5\arcsec to 3.2\arcsec (90\% of confidence). 
We corrected all the positions of the afterglows  observed by XRT in the first 11 months of the {\it Swift} mission. 
This is particularly important for the 37 X--ray afterglows without optical counterpart.
Optical follow-up of dark GRBs, in fact, will be more efficient with the use of the more accurate XRT positions.}
\maketitle
\section{Introduction}
The {\it Swift} satellite (Geherels et al 2004) detects and localises gamma ray bursts (GRBs) and provides autonomous 
rapid response observations and long term monitoring of their afterglow emission.
The scientific payload consists of three instruments: the Burst Alert Telescope (BAT, Barthelmy et al. 2005), the X--ray Telescope
(XRT, Burrows et al. 2005) and the UV/Optical Telescope (UVOT, Roming et al. 2005).
{\it Swift} observations provide prompt $\gamma$--ray positions with an accuracy of few arcminutes, 
X--ray positions  with an accuracy of few arcseconds and UV/Optical positions with accuracy of less than 1\arcsec.
From the satellite launch (2004 Nov 20) to 2005 Oct 16, XRT observed 64 afterglows of GRBs detected by BAT
and 5 afterglows of GRBs detected by other instruments (INTEGRAL and HETE-2). 

For the same GRB sample only 31 optical counterparts were found either by UVOT or 
by ground--based telescopes. In most of the remaining cases stringent upper limits were set by deep optical observations.
Classical explanations for burst optical darkness are, 
dust extinction, high redshift or intrinsic faintness. 
The study of the host galaxies through ultra--deep follow--up observations, 
is one of the most important ingredients for the determination of GRB progenitors. 
One of the key factors in the follow--up is the accuracy of the GRB position.
Before {\it Swift}, the study of the host galaxy was possible only in the presence of an optical afterglow detection
(with the exception for the rare cases observed by XMM--Newton, {\it Chandra} or by radio telescopes), because of the 
limited position accuracy of the X--ray and $\gamma$--ray telescopes.
This limitation  could have produced some bias in the statistical study of the population of the host galaxies.
For example, if we assume that the dust extinction plays an important role in the optical obscuration  
of the afterglow emission, the optically selected sample of host galaxies would have been biased against 
dust--rich galaxies. Now, by means of the {\it Swift} prompt automatic observations and the high quality of the XRT 
optics, we can obtain X--ray afterglow positions so accurate that the identification of the host galaxies is possible 
even when only the X--ray afterglow position is available.
One of the goals of the XRT is to provide afterglow X--ray positions with the unprecedented accuracy of 5\arcsec. 
From a comparison of the XRT positions with the optical transients positions we find that, at present, 90\%
of the XRT positions are within a 6.5\arcsec radius error circle, slightly worse than the pre--flight expectations.
While the statistical uncertainties in the XRT position determination are well known, we find that the larger 
fraction of the error has a  systematic origin.  
In the present work we show that we can correct for this systematic error, improving the typical accuracy, 
and, using the proposed correction, we recalculated 68 X--ray afterglow positions.
\section{The XRT position accuracy}
\begin{figure}[t]	
\includegraphics[width=9cm]{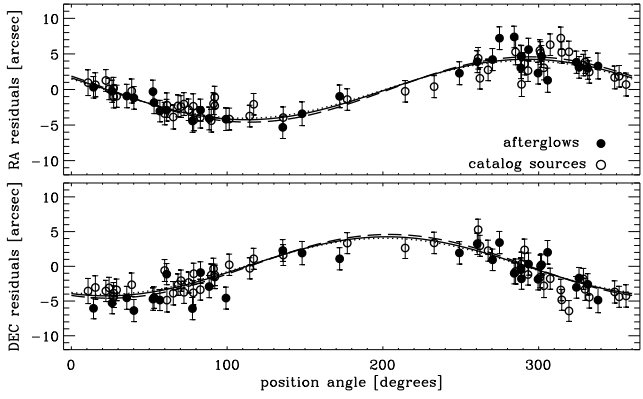}
\caption{Differences in coordinates (RA in the upper panel, DEC in the lower) between the XRT and optical
positions (X-optical). We used both catalogued sources and GRB observations with well known optical counterpart. 
The plotted errors are the combination of the statistical errors and the aspect solution uncertainty.
The solid line is the fit to the entire sample, the dashed line is the fit relative to the
catalogued sources only and the dotted line is the fit to the afterglows: the best fit parameters are perfectly 
consistent.   
RA residuals have been corrected by the factor cos(DEC), and represent the true separation in the sky.}
\label{fig:f2} 
\end{figure}
For each GRB, XRT usually produces two position measurements: the on--flight rapid position and the refined 
on--ground position.
The way the satellite works is that when the XRT observations starts promptly after the BAT trigger
(100-200 s) and the X--ray flux is bright enough 
($\sim$ 10 counts per second, corresponding to $\sim 5 \times 10 ^{-10}$ erg sec$^{-1}$ cm$^{-2}$), 
the first position measurement is calculated on--board by  the flight software (Hill et al. 2004) and automatically distributed 
by the GRB Coordinate Network (GCN), typically within a few seconds of the spacecraft slewing and settling on the GRB. 
The XRT was able to calculate and distribute a rapid position, within 350 seconds from the burst trigger,
in 19 cases. 
All the measured XRT positions are then refined on the ground using all the telemetered photon 
counting (PC) data from the first segment of the observation and sent again in a new GCN circular
(see Hill et al. 2005 for a detailed description of the XRT automatic procedure).
The uncertainties in the position measurements are determined by the statistical uncertainty and by the
precision of the satellite pointing.

When the background rate is negligible,
the statistical error in the position determination depends on the instrumental point spread function 
(PSF) and on the counts of the source according to the simple formula
U$_{\rm{stat}}\propto$R$_{90}/\sqrt{\rm counts}$, where U$_{\rm{stat}}$ is 90\%
accuracy error circle radius and R$_{90}$ is the radius which contains 90\% of the fluence.
The XRT statistical positional accuracy has been extensively tested on--ground and verified in flight during the 
calibration phase with some ad hoc observations (Hill et al. 2005; Moretti et al. 2005).
We empirically found that the statistical error at the 90\% level of XRT position measurement is given by  
the formula U$_{\rm{stat}}$=R$\times$counts$^{-0.48}$ (Hill et al. 2004), with the parameter R=23\arcsec,
in very good agreement with the expectations. 
It means that for a source with more than 144 counts, the statistical error is less than 2\arcsec.
To determine the afterglow position, the flight software uses a very short exposure image (0.1 or 2.5 seconds): 
this means that the typical source counts to determine a centroid with are $<$ 50
and therefore such positions are highly affected by statistical uncertainties. 
In telemetered PC mode data there are usually more than 150 counts and the statistical uncertainties in the 
on--ground refined position measurements are less than 2\arcsec.
We note that the expected background counts in a region containing 80\% of the PSF in a typical exposure time of 20 ks, in 
the full energy band (0.2-10 keV), are $\sim 2$ and this fully justifies the previous assumption that the 
background is negligible.

In addition to the statistical error, the XRT position uncertainty is also determined by the 
uncertainty in the satellite aspect solution. The nominal value of this uncertainty is 3\arcsec 
as reported in the calibration file swxposerr20010101v002.fits (CALDB version 20050916).

In order to study the systematic errors, we collected all the XRT observations of point--like sources 
present in the public archive (http://heasarc.gsfc.nasa.gov/cgi-bin/W3Browse/swift.pl) from April to 
August 2005 with a catalogued optical position in the SIMBAD archive (http://simbad.u-strasbg.fr/Simbad), 
excluding the few with proper motion. Moreover, we added the 15 calibration observations of Mkn 876, 
RXJ0720.4-3125 and RXSJ1708-4009 taken between February 2005 and March 2005. 
We then considered  all the observations of X--ray afterglows with a clearly varying optical counterpart published 
in the GCN in the period from December 2004 to September 2005. 
In order to select a very homogeneous sample and to minimize the statistical error, we selected the observations with 
the source at less than 3\arcmin ~ from the center of the field of view and with more than 150 source counts.
It resulted in a sample of 80 observations (31 afterglows and 49 catalogued sources).
All the XRT data were reduced using the {\it xrtpipeline} task of the current release of the HEADAS software 
(version 1.6), with all the default options and the current release of the calibration files (CALDB version 20050916); 
then we calculated the XRT positions by means of the {\it xrtcentroid} task .
\begin{figure}[t]	
\includegraphics[width=9cm]{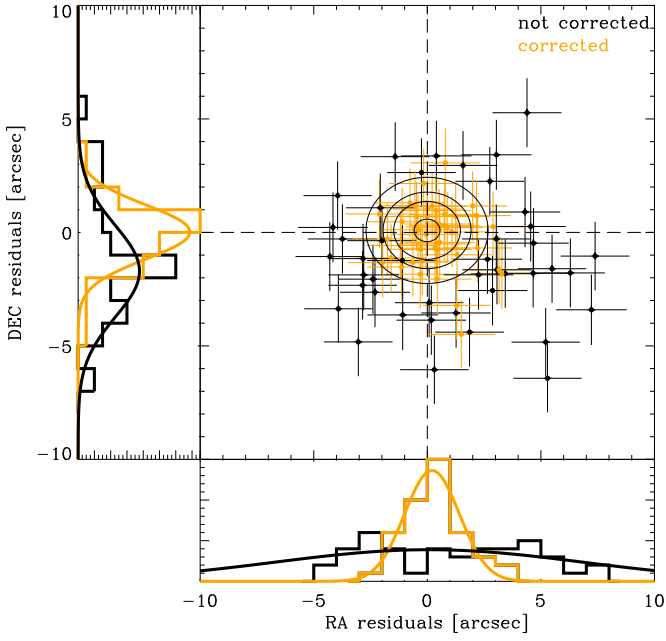}
\caption{The residuals of the 43 sources with more than 1000 counts before (black) and after 
the correction (grey). The contour of the two dimensional Gaussian best fit function are
over-plotted in correspondence of 10\%, 50\%, 90\% of the maximum value.
The external panels show the distribution of the coordinate residuals
before and after the correction together with their best Gaussian fit.}
\label{fig:f1} 
\end{figure}
As shown in Fig. \ref{fig:f2}, we found a clear relationship between the spacecraft position angle 
(PA, also called roll angle) of the observations and the residuals in RA and DEC of the X--ray positions 
in respect to the optical positions (note that RA residuals have been corrected by the factor cos (DEC), 
and represent the true separation in the sky).
The roll angle of an observation is the angle from sky coordinates to spacecraft coordinates
and it is available in the header keyword PA\_PNT of the event files.

We found that this relation is well fitted by a trigonometric
function. A similar relation was already found for the MECS and LECS
telescopes on board of the BeppoSAX satellite (Perri \& Capalbi
2002). This effect is due to a small calibration error between the XRT
boresight and the satellite star tracker boresight, which causes a
displacement of the detector system coordinate. The projection of the
this displacement in sky coordinates gives a dependence of the
coordinates residuals with the roll angle.  Therefore, we expect that
we can fit together the two relations with the following functions
\begin{equation}
\rm {\Delta(RA)= M \sin(PA+\phi)}  
\label{eq1}
\end{equation}
\begin{equation}
\rm {\Delta(DEC)=  M \cos(PA+\phi)} 
\label{eq2}
\end{equation}
where M is the amplitude of the misalignment and the phase $\phi$ is its direction. 
The best fit of the data is given by M=4.2\arcsec$\pm0.4$\arcsec~ and $\phi=157.6^{\circ} \pm 6.5^{\circ}$
(90\% confidence errors), $\chi^2$=0.8 for 158 degrees of freedom.
It corresponds to a shift of the nominal detector center of 1.8($\pm 0.2$) pixels  
(1 pixel is 2.36\arcsec). A refinement of the boresight calibration will be included in the 
standard calibration files from Dec 2005 CALDB distribution.
\begin{figure}[t]	
\includegraphics[width=9cm]{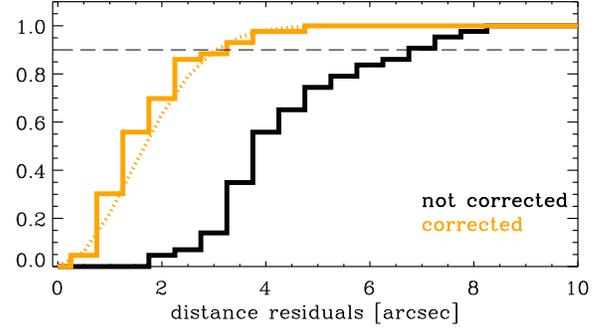}
\caption{The integral distribution of the distances of the X--ray sources, with more than 1000 counts, from the
optical counterpart before (black lines) and after the correction (grey lines). The dotted line is the 
the integral of the best Gaussian fit to the distribution of the coordinate residuals (see Fig.\ref{fig:f1}).}
\label{fig:f3} 
\end{figure}
Because, as explained below, we will use this fit to correct all the afterglow positions, 
we checked whether the two subsamples (the catalogued sources and the afterglows) yield consistent results.
As shown in Fig. \ref{fig:f2}, the fit of the two different sub--samples, as expected, give perfectly 
consistent results.  
\section{Correction of the systematic error}
The parametrisation of the relationship between the coordinate residuals and the roll angle of the observations
(Eqs. \ref{eq1}, \ref{eq2}) allows us to correct the source position derived from XRT observations with the 
following:
\begin{equation}
{\rm RA_{new}=  RA_{old}- \Delta(RA)/cos(DEC)}  
\label{eq4}
\end{equation}
\begin{equation}
{\rm DEC_{new}= DEC_{old}- \Delta(DEC)~.} 
\label{eq3}
\end{equation}
First, we tested the goodness of the correction on the 43 sources in the sample with more then 1000 counts
(statistical error $<$1\arcsec, Fig. \ref{fig:f1}). For all these 43 elements we calculated the distance between 
the optical position and the refined X--ray positions and we compared it with the uncorrected positions. 
Because for this particular source sub--sample the statistical uncertainty is negligible, this gives us the measurement 
of the systematic error.

In Fig. \ref{fig:f3} we compare the distribution of the distances before and after the correction.
From the integral distribution of the distances we find that 90\% of 
the X--ray sources after the correction were within a distance of 3.2\arcsec (while the uncorrected value 
was 6.5\arcsec). This is an improvement by a factor of $\sim$4 in terms of error circle area.
The mean of the distribution changes from 4.3\arcsec to 1.7\arcsec after the correction. 
We note that, as expected, the corrected residual distributions
are well consistent with a Gaussian distribution.  The Gaussian fit to
the RA residual distribution yields --0.03$\pm$0.18,
1.2$\pm$0.2 for the mean and standard deviation, respectively
($\chi^2=0.72$). The Gaussian fit to the DEC residual distribution
yields 0.13$\pm$0.20, 1.3$\pm$0.2 for the mean
and standard deviation, respectively ($\chi^2=0.60$).

We then applied the boresight correction to the whole sample of 68
afterglows observed by XRT in PC mode from the launch to 2005 Oct 16.
We retrieved all the first segments of the 68 PC observations
from the {\it Swift} archive.
We calculated the XRT positions by means of the {\it xrtcentroid} task. 
This task calculates the source centroids by recursively evaluating the 
barycentre in boxes reduced by 80\% each time from the initial box size, which
is an input parameter. 
If the statistical significance of the source is very low 
a too large detection box could affect the position
determination. Possible sources of error are a background fluctuations or 
faint serendipitous sources present in the centroid box.
In order to properly take into account for these effects we used the following procedure. 
First, we calculated the centroid with a fixed input error box size (30\arcsec);
then we refined this position using different box sizes as function of the
source counts, ranging from 30\arcsec (for bright sources) to 12\arcsec (for faint sources): 
for each source we used an initial box size such that the ratio between 
source and background counts, within the box, is always greater than 20, taking into 
account both the background rate and the PSF profile (Moretti et al. 2005).
Finally we applied the boresight corrections (Eqs. \ref{eq4},\ref{eq3}) 
to the so calculated positions.

The refined positions are reported in Table \ref{tab:grb}; the quoted errors (90\% confidence) 
are the quadratic sum 
of the statistical error with the new 3.2\arcsec systematic error. 
Moreover in the final error budget we added a term which depends on the source counts and which takes into account 
the uncertainty caused by the choice of the size of the xrtcentroid error box. We stress, however, that 
this term is negligible in most of the cases.
The mean error of the sample is 3.7\arcsec, and 90\% of the afterglows have 90\% confidence uncertainty less than 5\arcsec.  
We stress that among the 68 new positions, 37 are of dark GRBs. We excluded GRB 050117 from our analysis because 
it does not have any useful PC mode data. 
\begin{figure}	
\includegraphics[width=9cm]{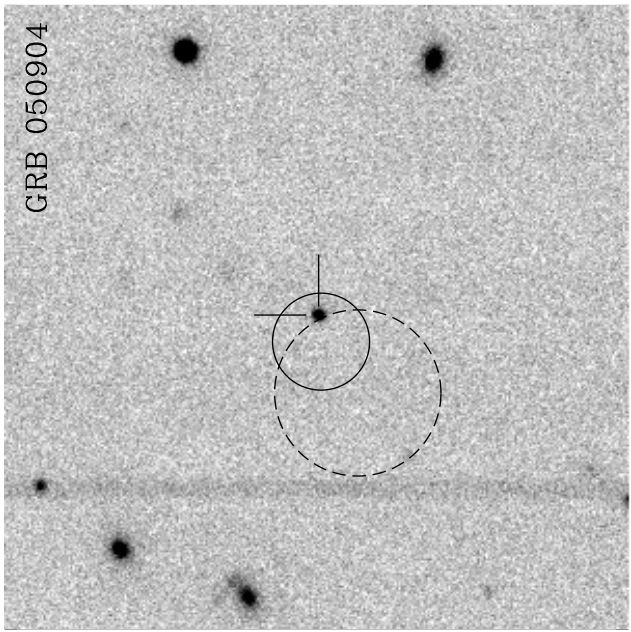}  
\caption{A VLT+ISAAC J filter image of GRB 050904. The lines show the position of the optical afterglow.
The dashed circle is the original error circle, while the smaller and solid line circle is the refined one. 
The optical transient is at 1.6\arcsec~from the refined XRT position and at
6.2\arcsec~from the original position quoted in Mineo et al. (2005).  
As it is clear from the distribution of the residuals (Fig. \ref{fig:f3}),
this represents an extreme case: only 15\% of the positions are expected 
to be at more than 6\arcsec from the optical position before the correction.
The image is taken from Tagliaferri et al. (2005).}
\label{fig:f4} 
\end{figure}

As an example of our results, in Fig. \ref{fig:f4} we illustrate the
case of GRB 050904 (Cummings et al. 2005), at redshift z=6.3 
(Kawai et al. 2005). 
In this case the prompt observations from the ground allowed the
detection of the IR transient (D'Avanzo et al. 2005). 
As shown in Fig. \ref{fig:f4} the IR source position is right at the border of the original XRT
error circle (6\arcsec radius, Mineo et al. 2005), while it is perfectly contained by the refined one
(3.2\arcsec).  
The measured colours of this afterglow do not show any hint of dust
extinction (Tagliaferri et al. 2005).
Had GRB 050904 been highly extincted, it probably would have been undetectable
by optical or IR observations. The search of the host galaxy 
would have been very hard because, although very close, the correct position
at the border of the uncorrected XRT error circle.

\section{Conclusion}
We refined the boresight calibration of the XRT, resulting in a significant improvement of the position accuracy.
The comparison with optical counterparts shows that on average we reduced the distance between XRT positions and 
optical positions by a factor of 2.
This was possible by reducing the systematic error from 6.5\arcsec to 3.2\arcsec  (90\% confidence).
By means of this correction we recalculated the position of 68 afterglows observed by XRT 
(the complete sample up to 2005 Oct 16). 
With one example, we showed how a deep follow--up study of the host galaxies of optically dark GRB  
can be performed with greater efficiency employing this correction.
We note that after Oct 16 2005 the XRT afterglow positions provided by the XRT team in the GCNs are 
calculated with the new boresight calibration. 
Therefore, this work provides the complete catalogue of the refined XRT afterglow positions of the GRB 
preceding that date. 
For the subsequent observations of GRB afterglows, we stress that a refined boresight calibration 
will be implemented in the standard calibration products from the Dec 2005 CALDB distribution. 
\begin{table*}
\begin{center}
\caption[]{Refined positions and 90\% statistical and systematic position
error radii for the whole sample of 68 GRB afterglows observed by the
Swift XRT.}
\footnotesize{
\begin{tabular}{|l|c|c|c||l|c|c|c|}
\hline
GRB & RA(J2000) & DEC(J2000) & unc.(90\%)[\arcsec] & GRB & RA(J2000) & DEC(J2000) & unc(90\%)[\arcsec]  \\
\hline
    041223 & 06 40 47.43 & --37 04  25.2 &   3.4 &    050713B & 20 31 15.51 &  +60 56  44.6 &   3.3 \\
    050124 & 12 51 30.33 &  +13 02  42.8 &   3.5 &     050714 & 02 54 22.93 &  +69 06  46.3 &   5.1 \\
    050126 & 18 32 27.13 &  +42 22  14.7 &   3.5 &    050714B & 11 18 47.66 & --15 32  48.9 &   3.3 \\
    050128 & 14 38 17.78 & --34 45  52.6 &   3.2 &     050716 & 22 34 20.77 &  +38 41  03.0 &   3.3 \\
   050215B & 11 37 47.70 &  +40 47  47.1 &   4.5 &     050717 & 14 17 24.60 & --50 32  00.0 &   3.3 \\
    050219 & 11 05 39.03 & --40 41  00.8 &   4.1 &     050721 & 16 53 44.62 & --28 22  52.1 &   3.3 \\
   050219B & 05 25 15.87 & --57 45  29.9 &   3.4 &     050724 & 16 24 44.64 & --27 32  25.3 &   3.4 \\
    050223 & 18 05 33.08 & --62 28  20.5 &   5.4 &     050726 & 13 20 11.95 & --32 03  50.6 &   3.3 \\
    050306 & 18 49 14.48 & --09 09  10.0 &   4.9 &     050730 & 14 08 17.22 & --03 46  18.8 &   3.2 \\
    050315 & 20 25 54.13 & --42 35  59.8 &   3.2 &     050801 & 13 36 35.37 & --21 55  42.1 &   3.4 \\
    050318 & 03 18 50.77 & --46 23  44.8 &   3.3 &     050802 & 14 37 05.84 &  +27 47  10.8 &   3.2 \\
    050319 & 10 16 47.80 &  +43 32  54.9 &   3.2 &     050803 & 23 22 37.90 &  +05 47  08.8 &   3.2 \\
    050326 & 00 27 49.18 & --71 22  14.6 &   3.6 &     050813 & 16 07 57.07 &  +11 14  54.2 &   6.5 \\
    050401 & 16 31 28.85 &  +02 11  14.4 &   3.3 &     050814 & 17 36 45.43 &  +46 20  22.3 &   3.3 \\
    050406 & 02 17 52.39 & --50 11  14.9 &   3.8 &     050815 & 19 34 22.94 &  +09 08  50.8 &   3.5 \\
    050408 & 12 02 17.35 &  +10 51  09.6 &   3.3 &     050819 & 23 55 01.45 &  +24 51  35.3 &   3.7 \\
    050410 & 05 59 12.74 &  +79 36  09.8 &   4.0 &     050820 & 22 29 38.16 &  +19 33  35.1 &   3.2 \\
    050412 & 12 04 25.19 & --01 12  00.4 &   4.2 &    050820B & 09 02 25.48 & --72 38  43.3 &   4.1 \\
    050416 & 12 33 54.63 &  +21 03  27.3 &   3.3 &     050822 & 03 24 27.09 & --46 01  59.6 &   3.3 \\
    050421 & 20 29 03.18 &  +73 39  18.2 &   3.5 &     050824 & 00 48 56.23 &  +22 36  31.2 &   3.4 \\
    050422 & 21 37 54.92 &  +55 46  45.3 &   3.7 &     050826 & 05 51 01.49 & --02 38  38.6 &   3.4 \\
   050502B & 09 30 10.01 &  +16 59  47.1 &   3.3 &     050827 & 04 17 09.58 &  +18 12  00.2 &   3.4 \\
    050504 & 13 24 01.19 &  +40 42  15.7 &   5.3 &     050904 & 00 54 50.82 &  +14 05  08.2 &   3.2 \\
    050505 & 09 27 03.19 &  +30 16  22.7 &   3.2 &     050908 & 01 21 50.65 & --12 57  19.0 &   3.4 \\
    050509 & 20 42 19.86 &  +54 04  16.3 &   4.6 &     050915 & 05 26 44.81 & --28 01  00.3 &   3.4 \\
   050509B & 12 36 13.56 &  +28 59  01.7 &   7.6 &    050915B & 14 36 26.26 & --67 24  32.1 &   3.4 \\
    050520 & 12 50 05.77 &  +30 27  03.7 &   4.7 &     050916 & 09 03 56.98 & --51 25  46.6 &   3.3 \\
    050522 & 13 20 34.63 &  +24 47  20.4 &   9.2 &    050922B & 00 23 13.23 & --05 36  17.2 &   3.2 \\
    050525 & 18 32 32.63 &  +26 20  21.5 &   3.4 &    050922C & 21 09 33.12 & --08 45  29.6 &   3.3 \\
    050603 & 02 39 56.82 & --25 10  55.2 &   3.5 &     051001 & 23 23 48.77 & --31 31  20.9 &   3.5 \\
    050607 & 20 00 42.78 &  +09 08  30.5 &   3.3 &     051006 & 07 23 14.02 &  +09 30  19.5 &   3.4 \\
    050701 & 15 09 01.67 & --59 24  53.7 &   3.7 &     051008 & 13 31 29.50 &  +42 05  55.7 &   3.2 \\
    050712 & 05 10 48.00 &  +64 54  48.0 &   3.3 &     051016 & 08 11 16.65 & --18 17  55.1 &   3.5 \\
    050713 & 21 22 09.78 &  +77 04  28.9 &   3.3 &    051016B & 08 48 27.70 &  +13 39  18.8 &   3.3 \\
\hline
\end{tabular}
}
\label{tab:grb}  
\end{center}
\end{table*}

\begin{acknowledgements}
This work is supported at OAB--INAF by ASI grant I/R/039/04, at Penn State by NASA contract NAS5-00136 and
at the University of Leicester by PPARC of grant PPA/Z/S/2003/00507.
This research has made use of the SIMBAD database, operated at CDS, Strasbourg, France.
\end{acknowledgements}


\end{document}